\documentclass{pasj00}

\begin{document}
\SetRunningHead{Yasumi et al.}
{Suzaku Observations of G350.1$-$0.3 and G349.7$+$0.2.}
\Received{2013/11/12}
\Accepted{2014/03/19}

\title{Abundances in the Ejecta of Core Collapse Supernova Remnants, G350.1$-$0.3 and G349.7$+$0.2}


 \author{Masato \textsc{Yasumi}\altaffilmark{1},
   Masayoshi \textsc{Nobukawa}\altaffilmark{2,1},
   Shinya \textsc{Nakashima}\altaffilmark{1},
   Hiroyuki \textsc{Uchida}\altaffilmark{1},
   Ryusuke \textsc{Sugawara}\altaffilmark{1},\\
   Takeshi Go \textsc{Tsuru}\altaffilmark{1},
   Takaaki \textsc{Tanaka}\altaffilmark{1},
   and 
Katsuji \textsc{Koyama}\altaffilmark{1,3}}  
 \altaffiltext{1}{Department of Physics, Graduate School of Science, Kyoto University,
  Kitashirakawa Oiwake-cho, Sakyo-ku, Kyoto 606-8502, Japan}
 \email{yasumi@cr.scphys.kyoto-u.ac.jp, nobukawa@cr.scphys.kyoto-u.ac.jp}
 \altaffiltext{2}{The Hakubi Center for Advanced Research, Kyoto University,
  Yoshida-Ushinomiya-cho, Kyoto 606-8302, Japan}
\altaffiltext{3}{Department of Earth and Space Science, Graduate School of Science,
Osaka University, 1-1 Machikaneyama, Toyonaka, Osaka 560-0043, Japan}

\KeyWords{X-rays: individuals: G350.1$-$0.3 --- X-rays: individuals: G349.7$+$0.2 --- X-rays: ISM} 

\maketitle

\begin{abstract}

We present Suzaku results of the two Galactic supernova remnants (SNRs), G350.1$-$0.3 and G349.7$+$0.2.
We find Al and Ni K$\alpha$ lines from both the SNRs for the first time, in addition to previously detected K-shell lines of Mg, Si, S, Ar, Ca and Fe.  
The spectra are well described by two optically thin thermal plasmas: a low-temperature (low-$kT$) plasma in collisional ionization equilibrium and a high-temperature (high-$kT$) plasma in non-equilibrium ionization.
Since the low-$kT$ plasma has solar metal abundances, it is thought to be of interstellar medium origin.
The high-$kT$ plasma has super-solar abundances, hence it is likely to be of ejecta origin.
The abundance patterns of the ejecta components are similar to those of core-collapse supernovae with the progenitor mass of $\sim$ 15--25~\MO\ for G350.1$-$0.3 and $\sim$ 35--40~\MO\ for G349.7$+$0.2.
We find extremely high abundances of Ni compared to Fe ($Z_{\rm Ni}/Z_{\rm Fe} \sim8$). 
Based on the measured column densities between the SNRs and the near sky background, 
we propose that G350.1$-$0.3 and G349.7$+$0.2 are located at the distance of $9\pm3$~kpc and $12\pm5$~kpc, respectively.
Then the ejecta masses are estimated to be $\sim$13 \MO\ and $\sim$24~\MO\ for G350.1$-$0.3 and G349.7$+$0.2, respectively.
These values are consistent with the progenitor mass of $\sim$ 15--25~\MO\ and $\sim$ 35--40~\MO\ for G350.1$-$0.3 and G349.7$+$0.2, respectively.

\end{abstract}

\section{Introduction}
G350.1$-$0.3 is a radio-bright supernova remnant (SNR) in the Galaxy \citep{Clark1973}.
The radio morphology is not a typical shell or crab-like but has a distorted and elongated shape \citep{Salter1986}.
X-rays were detected with ROSAT \citep{Voges1999} and ASCA \citep{Sugizaki2001}, and then with XMM-Newton \citep{Gaensler2008} and Chandra \citep{Lovchinsky2011}.
A point-like X-ray source, XMMU\,J172054.5$-$372652, was found $\sim\timeform{3'}$ west of the brightest region.
The spectral parameters are in the range of a typical central compact object (CCO) \citep{Lovchinsky2011}.
\citet{Gaensler2008} detected a $^{12}$CO emission along the eastern edge of the SNR. 
They suggest that the molecular gas suppressed the expansion of the remnant and formed the peculiar asymmetric morphology.
The distance and age are estimated to be 4.5--10.7~kpc and $\sim900$ years old, respectively \citep{Gaensler2008}.
The X-ray spectrum of the SNR is reproduced by a two-component model: a high-temperature ($kT\sim1.5$~keV) plasma in non-equilibrium ionization (NEI) and a low-temperature ($kT\sim0.4$~keV) plasma in collisional ionization equilibrium (CIE) \citep{Gaensler2008}.
The abundances of the former are super-solar and those of the latter are 1~solar, which suggests ejecta and interstellar medium (ISM) origin, respectively.
The metal abundances of the ejecta are as high as 10~solar although the statistical errors are quite large.
\citet{Lovchinsky2011}, on the other hand, reported that the spatially resolved spectra can be reproduced by one-temperature plasma models with abundances of $\sim1$--9~solar.

G349.7$+$0.2 is another radio-bright SNR in the Galaxy \citep{Shaver1985}.
The distance is estimated to be $18.3\pm 4.6$~kpc \citep{Caswell1975}.
OH maser emission (1720 MHz) is found toward G349.7$+$0.2 at a radial velocity of $\sim+16$~km~s$^{-1}$, suggesting that the SNR is interacting with a dense molecular cloud at the kinematic distance of 22.4~kpc \citep{Frail1996}. 
The X-ray image taken by Chandra shows an irregular shell with the bright eastern side \citep{Lazendic2005}.
The presence of H\emissiontype{I} clouds near the SNR indicates that G349.7$+$0.2 is evolved into the intercloud medium, and is responsible for the irregular morphology.
Like G350.1$-$0.3, the X-ray spectrum is described by two plasmas with different temperatures: a low-temperature ($kT\sim0.8$~keV) CIE plasma with solar abundances, and a high-temperature ($kT\sim1.4$~keV) NEI plasma. 
The latter has an enhanced Si abundance, suggesting ejecta origin \citep{Lazendic2005}.
A point source, CXOU\,J171801.0$-$372617, is found near the center of the SNR, possibly a CCO associated with the SNR \citep{Lazendic2005}.

The previous results such as the presence of CCOs and the associations with molecular clouds suggest that both of G350.1$-$0.3 and G349.7+0.2 are core-collapse (CC) SNRs.
The metal abundances in the ejecta should provide crucial information for the mass of the progenitor stars.
Fe and Ni are the final products of the major nuclear reaction network in the evolution of massive stars and their final supernova (SN) explosions.
Therefore, these elements should be particularly important to study the mechanism of CC SNe in the close vicinity of the core region.
Previous works, however, have limited statistics to study the ejecta elements.
This paper presents the most accurate Fe and Ni abundances in the two CC-SN candidates, G350.1$+$0.3 and G349.7$-$0.2.
For the studies, we used the Suzaku satellite \citep{Mitsuda2007} because it has the highest sensitivity for diffuse X-rays in the Fe and Ni K-shell band at 5--10~keV.
 
In this paper, we estimate errors at 90\% confidence level while figure~2, 3 and 6 show the 1~$\sigma$ errors. 
 
\section{Observations and Data Reduction}

The observations of G350.1$-$0.3 and G349.7$+$0.2 were made with the X-ray Imaging Spectrometer (XIS: \cite{Koyama2007}) on the focal planes of X-ray telescopes (XRT: \cite{Serlemitsos2007}) onboard the Suzaku satellite.
The observation log is given in table~\ref{tab:obs}.
The effective exposure times are 70.1~ks and 160.4~ks for G350.1$-$0.3 and G349.7$+$0.2, respectively.
We obtained cleaned event data after the pipeline processing version 2.7.16.30 from the Suzaku database.
We re-processed the data with the calibration database released in November 2012.

The XIS has four CCDs (XIS\,0, 1, 2, and 3). 
XIS\,0, 2, and 3 are Front-Illuminated (FI) CCDs and XIS\,1 is a Back-Illuminated (BI) CCD.
XIS\,2\footnote{http://www.astro.isas.ac.jp/suzaku/doc/suzakumemo/\\suzakumemo-2007-08.pdf} and one forth (Segment A) of XIS\,0\footnote{http://www.astro.isas.ac.jp/suzaku/doc/suzakumemo/\\suzakumemo-2010-01.pdf} have been out of function since November 2006 and June 2009, respectively.
The 1.70--1.76~keV band including the neutral Si K-shell edge is ignored because of the calibration uncertainty.
For data reprocess and analysis, we use the HEAsoft package version 6.11.

\begin{table*}[!t]
\caption{Observation log.}\label{tab:obs}
\begin{center}
\begin{tabular}{cccccc}
\hline              
 Name & Obs. ID & Obs. Date & (R.A., Dec.) $_{J2000}$ & Exposure\\
\hline              
 G350.1$-$0.3& 506065010  & 2011-Sep-17& (\timeform{260.2697D}, \timeform{-37.4549D}) & 70.1\ ks\\
 G349.7$+$0.2 & 506064010 & 2011-Sep-29& (\timeform{259.4954D}, \timeform{-37.4452D}) &160.4\ ks\\
 \hline
\end{tabular}
\end{center}
\end{table*}

\section{Analysis and Results}

\subsection{X-ray Images}

\begin{figure}[t!]
\begin{center}
\FigureFile(80mm,100mm){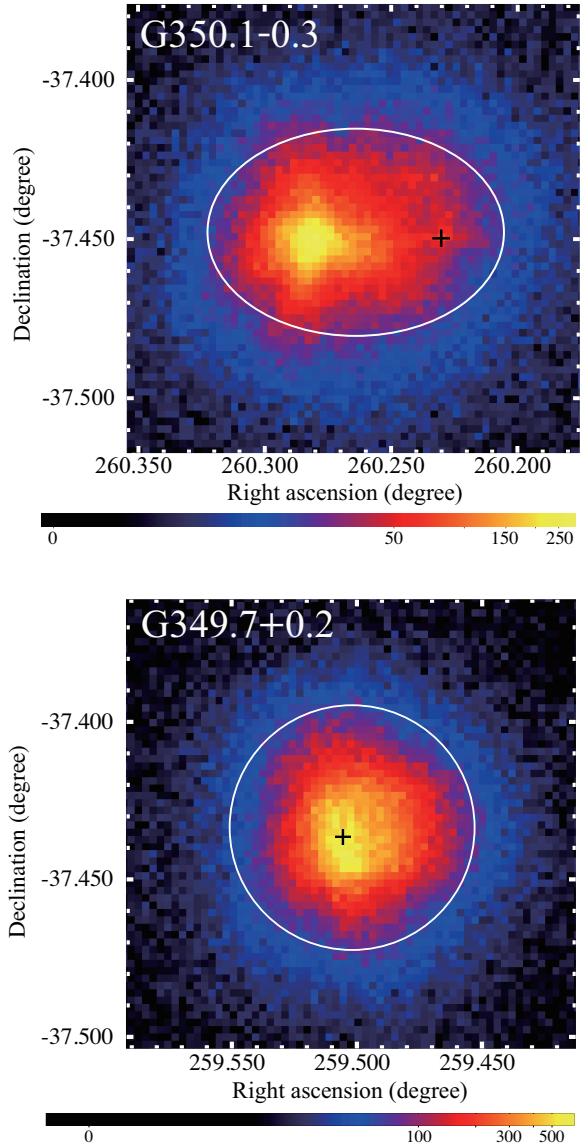}
\caption{NXB-subtracted images of $\timeform{8.5'}\times\timeform{8.5'}$ in the 1.0--10.0~keV band after the correction of the vignetting effect.
The data of XIS\,0, 1 and 3 are co-added.
Color scale shows X-ray counts in $\timeform{8.3"}\times\timeform{8.3"}$.
Source spectra of G350.1$-$0.3 are extracted from the white ellipse with semi-major and semi-minor axes of \timeform{2.8'} and \timeform{2.0'}, respectively, while those of G349.7$+$0.2 are taken from the white circle with a radius of $\timeform{2.3'}$. 
The cross marks are the positions of the CCOs, XMMU\,J172054.5$-$372652 \citep{Gaensler2008} and CXOU\,J171801.0$-$372617 \citep{Lazendic2005}, in G350.1$-$0.3 and G349.7$+$0.2, respectively.}
\label{img}
\end{center}
\end{figure}

We show the 1.0--10.0~keV band images in the $\timeform{8.5'}\times\timeform{8.5'}$ fields at the center of the two observations in figure~\ref{img}.
The non X-ray background (NXB) is made using {\tt xisnxbgen} \citep{Tawa2008}, and is subtracted from the raw images.
The emission of G350.1$-$0.3 consists of a bright clump in the east with a radius of $\sim\timeform{1.5'}$ and a fainter emission extending toward the west.
The emission of G349.7$+$0.2 is circular with a radius of $\sim\timeform{2'}$.
The positions of both the CCOs, XMMU\,J172054.5$-$372652 \citep{Gaensler2008} and CXOU\,J171801.0$-$372617 \citep{Lazendic2005}, are given by the cross marks.

\subsection{X-ray Spectra}

The NXB-subtracted spectra of the SNRs and the background (BG) are given in figure~\ref{src_bg}, with the black and gray data points, respectively.
We extract the SNR spectra of G350.1$-$0.3 and G349.7$+$0.2 from the source regions enclosed by the white ellipse and circle shown in figure~\ref{img}, respectively.
The BG spectra for each SNR are taken from surrounding regions in each field-of-view. 
In the following spectral analysis, we use the XSPEC software version 12.7.0 \citep{Arnaud1996}.
The redistribution matrix files and ancillary response files are generated by {\tt xisrmfgen} and {\tt xissimarfgen}, respectively \citep{Ishisaki2007}.
The abundances are referred to \citet{Anders1989}.

\subsubsection{Background Estimation}

\begin{figure}[t!]
\begin{center}
\FigureFile(80mm,130mm){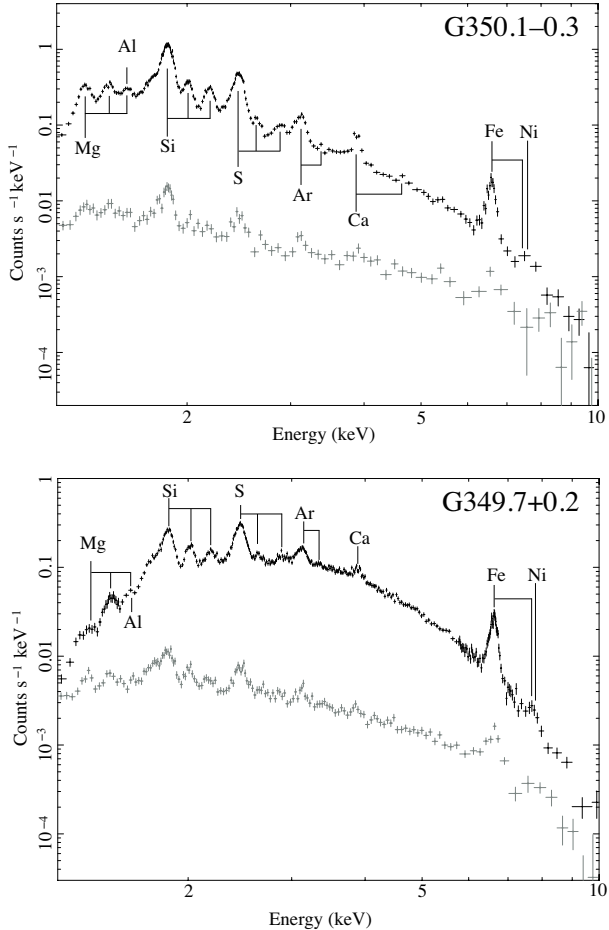}
\caption{NXB-subtracted spectra of G350.1$-$0.3 and G349.7$-$0.2. 
The data of FI CCDs (XIS\,0 and 3) are co-added. 
The spectra from the source and BG regions are shown in black and gray, respectively.}
\label{src_bg}
\end{center}
\end{figure}

The BG spectra in the energy band of Fe K-shell lines ($>$ 6~keV) have larger statistical errors than those of the SNR spectra (see figure~\ref{src_bg}), and hence a direct subtraction of the BG from the SNR spectra causes large statistical errors in this energy band.
We therefore make a BG model, then add it to the SNR spectra in the fitting procedures.
To make the BG model, we fit the BG spectra of XIS\,0, 1 and 3 simultaneously, allowing small offset energies in each XIS, because the absolute gain of the XIS has an uncertainty of $\sim5$~eV \citep{Koyama2007}.

Since both the SNRs are located near the inner Galactic disk, the BG is dominated by the Galactic ridge X-ray emission (GRXE) (e.g. \cite{Uchiyama2013}).
In fact, we see emission lines from Mg, Si, S, and Fe in the BG spectra as well as the hard continuum (figure~\ref{src_bg}), which are typical features in the GRXE spectra. 
\citet{Uchiyama2013} represented the GRXE spectra as the sum of high-temperature ($kT\sim7$~keV) and low-temperature ($kT\sim1$~keV) plasmas (here, HP and LP, respectively) in CIE ({\tt apec} model in XSPEC) plus non-thermal X-rays of the photon index $\Gamma \sim2$ with additional neutral Fe lines at 6.40~keV of the equivalent width ($EW$) $\sim460$~eV and at 7.06~keV. 
The latter component would be associated with cold matter (hence, CM) \citep{Uchiyama2013}.
The GRXE components are subject to a significant interstellar absorption ($N_{\rm H}$; {\tt phabs} model).
Thus, the model of GRXE is
\begin{equation}
{\rm GRXE}=N_{\rm H (GRXE)}\times({\rm LP + HP + CM}).
\end{equation}
In the fitting, the absorption column density $N_{\rm H (GRXE)}$ and the fluxes of HP, LP (emission measure: $EM$) and CM are free parameters.
The temperatures ($kT_{\rm LP}$ and $kT_{\rm HP}$) and abundances are also free, but are common between the two BG regions.

In addition to the GRXE, the Galactic plane background includes foreground thermal emissions (FE) (\cite{Ryu2009}; \cite{Uchiyama2013}).  
We fix the FE components according to \citet{Uchiyama2013}: absorbed two optically thin thermal plasmas ($N_{\rm H (FE)}=5.6\times10^{21}$~cm$^{-2}$, $kT=0.09$~keV and 0.59~keV).
The intensities in the 0.5--1.2~keV band are $1.1\times10^{-6}$~photons~s$^{-1}$~cm$^{-2}$~arcmin$^{-2}$ and $9.7\times10^{-7}$~photons~s$^{-1}$~cm$^{-2}$~arcmin$^{-2}$ for the 0.09~keV and 0.59~keV plasma, respectively.

We further add an absorbed power-law for the cosmic X-ray background (CXB).
The parameters of the CXB model are taken from \citet{Kushino2002}: $\Gamma =1.41$, and the flux of $6.38\times10^{-8}$~erg~s$^{-1}$~cm$^{-2}$~sr$^{-1}$.
The column density is assumed to be $N_{\rm H (CXB)}=2\times N_{\rm H (GRXE)}$ \citep{Uchiyama2013}.
The overall background model (BGD) is then given by
\begin{equation}
{\rm BGD = GRXE + FE + CXB}.
\end{equation}

Since G350.1$-$0.3 and G349.7$+$0.2 are almost in the same direction, we simultaneously fit the BG spectra for both the SNRs. The best-fit  parameters are listed in table~\ref{tab:bg}. 
We note that these parameters are globally consistent with \citet{Uchiyama2013}.
We hence apply the best-fit models for the X-ray BG of the SNRs. 

\begin{table*}[t!]
 \caption{Fitting results of the BG spectra\footnotemark[$*$].}\label{tab:bg}
  \begin{center}
    \begin{tabular}{llcc}
  \hline
   Component &Parameter& for G350.1$-$0.3& for G349.7$+$0.2 \\
      \hline

      Absorption & $N_{\rm H (GRXE)}$ ($\times10^{22}$~cm$^{-2}$) & $3.0\pm0.1$ & $3.9\pm0.1$ \\
     
      LP & $kT$ (keV)  &\multicolumn{2}{c}{$0.82\pm0.03$}\\
                             &  Abundance (solar)  & \multicolumn{2}{c}{$0.73\pm0.10$} \\
                             &  $EM$\footnotemark[$\dagger$] ($\times10^{12}$~cm$^{-5}$)&  $0.63\pm0.07$&  $1.6\pm0.3$\\
      HP & $kT$ (keV)  &\multicolumn{2}{c}{$5.1\pm0.6$}\\
                             &  Abundance (solar)  &  \multicolumn{2}{c}{(=LP)}\\
                             &  $EM$\footnotemark[$\dagger$] ($\times10^{11}$~cm$^{-5}$) &$0.78\pm0.10$&$1.2\pm0.1$\\
                             
     CM       & $\Gamma$ & \multicolumn{2}{c}{$2.13$ (fixed)} \\
                            &$EW_{\rm 6.40}$(eV)&\multicolumn{2}{c}{457~(fixed)}\\
                            &Flux\footnotemark[$\ddagger$] ($\times10^{-5}$~photons~s$^{-1}$~cm$^{-2}$)&  $< 4.0$ & $8.0\pm2.4$\\ 
                         \hline
      $\chi ^2$/d.o.f. &  & \multicolumn{2}{c}{$1051/769=1.37$}\\
      \hline
\multicolumn{3}{@{}l@{}}{\hbox to 0pt{\parbox{180mm}{\footnotesize
\par\noindent
\footnotemark[$*$]Errors are at the 90\% confidence level.
\par\noindent
\footnotemark[$\dagger$]  The emission measure in unit of $n_{\rm{e}}n_{\rm{H}}V/4\pi d^{2}$, where $n_{\rm{e}}$, $n_{\rm{H}}$, $V$ and $d$ are the electron and the hydrogen densities,\\
 the emitting volume and the distance to the source, respectively.
\par\noindent
\footnotemark[$\ddagger$] Absorbed flux (1.0--10.0~keV).}\hss}}
    \end{tabular}
     \end{center}
\end{table*}

\subsubsection{X-ray Spectra of the SNRs}\label{SNR_fit}

The spectra of G350.1$-$0.3 and G349.7$+$0.2 have stronger K$\alpha$ lines of Mg, Si, S, Ar, Ca and Fe than those in the BG spectra (figure~\ref{src_bg}).
The structure at 7.7~keV is a new discovery, which is composite emissions of Fe K$\beta$ and Ni K$\alpha$.
In addition, the structure at 1.6~keV is due either to He-like Mg K$\beta$ or Al K$\alpha$.
For the SNR spectral fitting (1.2--10.0~keV), we use the {\tt vapec} and {\tt vpshock} models in XSPEC for CIE and NEI plasmas, respectively, adding the background model given in table~\ref{tab:bg}.
 
\subsubsection*{G350.1$-$0.3}

\begin{figure}[t!]
\begin{center}
\FigureFile(80mm,150mm){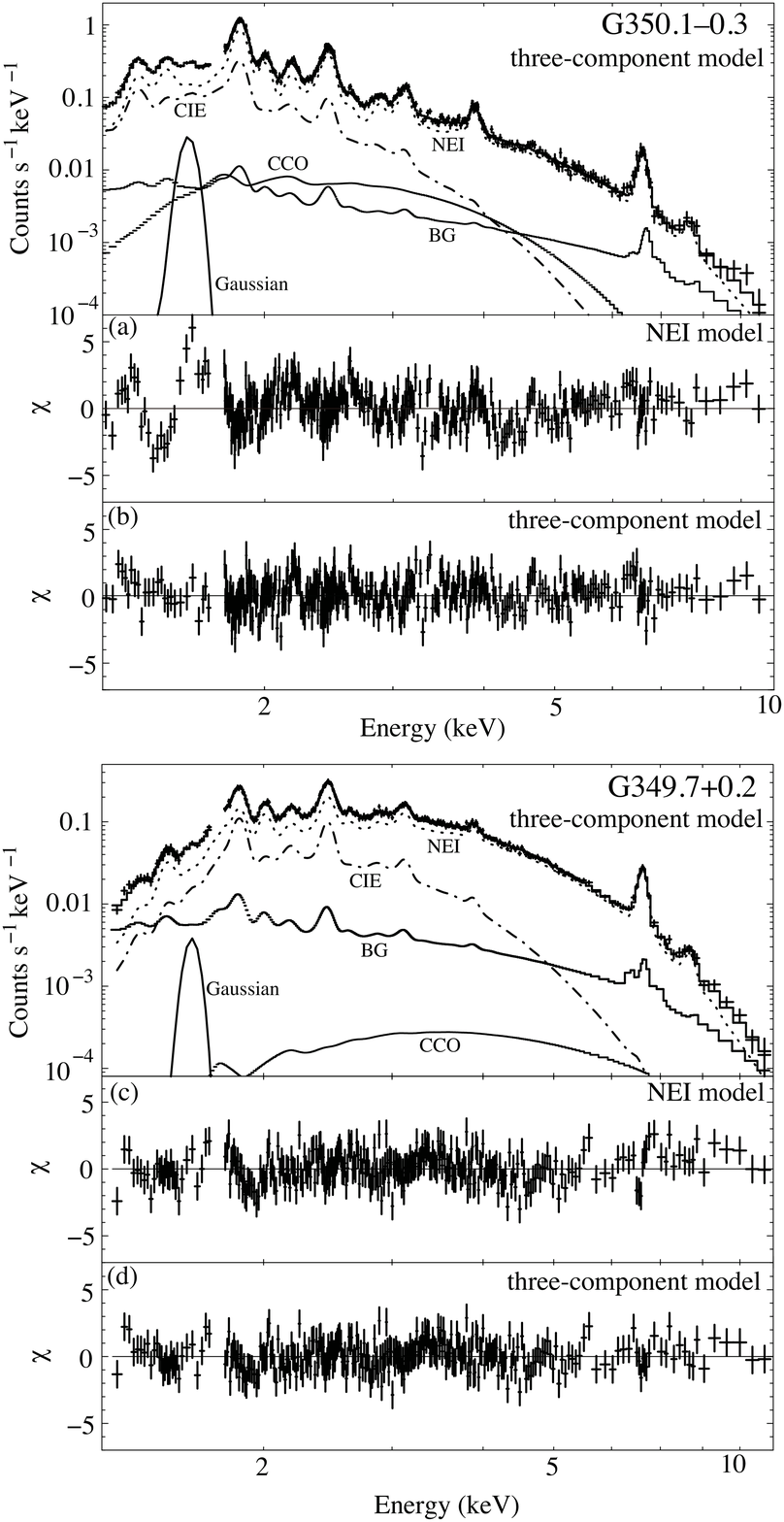}
\caption{G350.1$-$0.3 and G349.7$+$0.2 spectra extracted from the source regions.
Only the FI spectra are displayed for visibility.
The dash-dotted and dotted lines show the CIE and NEI models, respectively.
The solid lines show the BG, the CCO and the Gaussian models.
The lower panels (a) and (b) show the residuals from an NEI model and the three-component models of G350.1$-$0.3, respectively, while panels (c) and (d) are those for G349.7$+$0.2.}
\label{fig:fit}
\end{center}
\end{figure}

We first apply an NEI with interstellar absorption. 
The temperature ($kT$), ionization timescale ($n_{\rm e}t$), emission measure ($EM$) and abundances of Mg, Si, S, Ar, Ca, Fe and Ni are free parameters.
Those of the other elements are fixed to the solar values.
Since the source region includes the CCO candidate, XMMU\,J172054.5$-$372652, we add the spectral model for the CCO reported by \citet{Gaensler2008}: the blackbody ($kT=0.53$~keV, the 0.5--10.0~keV flux of $1.4\times 10^{-12}$~erg~s$^{-1}$~cm$^{-2}$) with $N_{\rm H}=2.9\times10^{22}$~cm$^{-2}$.
This model is not statistically acceptable ($\chi ^2/{\rm d.o.f.}=1279/663=1.92$), with significant residuals at $\sim1.35$~keV and $\sim1.45$~keV (see figure~\ref{fig:fit}a).
The residuals correspond to He-like and H-like Mg K$\alpha$, respectively, and hence at least one more plasma is required.
 
Then we add a CIE component with free parameters of Mg, Si, S, Ar, Ca, Fe and Ni abundances.
The other elements in the CIE component are fixed at the solar values.
The two-component model improve the fitting with $\chi^{2}/{\rm d.o.f.}=868/658=1.32$.
In the CIE component, although the abundances of Mg and Si are $\sim1$~solar, those of S, Ar, Ca, Fe and Ni are not constrained.
We, therefore, re-fit the spectrum fixing all the abundances of the CIE component at the solar values.
The model also gives a good fit ($\chi^{2}/{\rm d.o.f.}=888/661=1.34$), but still exhibits residuals at $\sim1.6$~keV. 
Then we add a narrow Gaussian to the model.
The three-component model yields $\chi^{2}$/d.o.f. of $873/659=1.32$.
The best-fit model and its parameters are given in figure~\ref{tab:para} and table~\ref{tab:para}, respectively.
There are the residuals around the Fe K$\alpha$ line.
We separately checked each XIS data and found that the residuals are only in XIS\,0.
These are due to a calibration error in the Fe K$\alpha$ line of the XIS\,0 data but do not affect the fitting results.

The center energy of the Gaussian line ($1.58\pm0.02$~keV) corresponds to the energies of He-like Mg K$\beta$ and Al K$\alpha$.
Mg K$\beta$ is included in both the CIE and NEI models while Al K$\alpha$ is included only in the CIE model.
Thus, the Gaussian line is Al K$\alpha$ from the NEI plasma.
The emissivity for the He-like Al K$\alpha$ is calculated to be $\varepsilon = 2.4\times10^{-11}$~cm$^{3}$~s$^{-1}$ for the best-fit temperature and ionization timescale (H. Yamaguchi, private communication).
Then the line intensity of $2.0(\pm0.7)\times10^{-4}$~photons~s$^{-1}$~cm$^{-2}$ is converted to an Al abundance of $1.4\pm0.5$~solar.
This result is also shown in table~\ref{tab:para}.

In table~\ref{tab:para}, we see an extreme overabundance of Ni for the NEI component ($14\pm7$~solar).
The Ni abundance is extracted from the peak flux at 7.7~keV, where the Fe K$\beta$ and Ni K$\alpha$ lines are not resolved.
In the plasma of the best-fit model ($kT=1.51$~keV, $n_{\rm e}t=3.5\times10^{11}$~s~cm$^{-3}$), the most populous Fe ions are Be, B and C-like states.
The present NEI model, however, does not include any K$\beta$ emission except for H-like and He-like states.
According to \citet{Yamaguchi2014}, we calculate the intensity ratio Fe K$\beta$/Fe K$\alpha$ and Fe K$\beta$ energies for each ion using ``Flexible Atomic Code'' \citep{Gu2008}.
The intensity ratio Fe K$\beta$/Fe K$\alpha$ is 2.9\%, which is in fact much larger than 0.2\%, the predicted value using the relevant NEI model.
The calculated mean energy of missing Fe K$\beta$ is 7.64~keV.
We, hence add a Gaussian at 7.64~keV with the intensity of $5.5\times10^{-7}$~photons~s$^{-1}$~cm$^{-2}$, 2.7\% of the Fe K$\alpha$ flux, and re-fit.
The additional Gaussian values and the re-fitted Ni abundance are added in table~\ref{tab:para}.
The parameters except for Ni are almost the same as those of the  previous three-component model (no correction of Fe K$\beta$).
In this revised three-component model, Ni abundance decreases from $14\pm7$~solar to $12\pm7$~solar.
Still, the overabundance of Ni does hold.
 
\begin{table*}[htbp]
 \caption{The best-fit parameters for source spectra.\footnotemark[$*$]}\label{tab:para}
  \begin{center}
    \begin{tabular}{llccc}
       \hline
          Component & \multicolumn{2}{l}{Parameter}&  G350.1$-$0.3 & G349.7$+$0.2\\
      \hline
      Absorption & $N\rm _H$ ($\times10^{22}$~cm$^{-2}$) && $3.3\pm0.1$ &$6.4\pm0.2$\\
     CIE &$kT$ (keV)  && $0.48\pm0.04$&$0.60\pm0.04$ \\
     &Abundance (solar)&& 1 (fixed)& 1 (fixed) \\
      &$EM$\footnotemark[$\dagger$] ($\times10^{13}$~cm$^{-5}$)&&$1.3\pm0.3$&$1.3\pm0.3$\\
      NEI &$kT$ (keV)  & &1.51$\pm$0.09&$1.24\pm0.03$ \\
      &Abundance (solar)&Mg&$3.7\pm0.5$&$3.6\pm1.1$\\
      &&Al\footnotemark[$\ddagger$]&$1.4\pm0.5$&$0.6\pm0.4$\\
      &&Si&$4.0\pm0.3$&$1.10\pm0.14$\\
      &&S&$2.8\pm0.2$&$0.72\pm0.04$\\
      &&Ar&$2.7\pm0.3$&$0.71\pm0.07$\\
      &&Ca&$3.7\pm0.4$&$0.67\pm0.10$\\
      &&Fe&$1.4\pm0.2$&$0.63\pm0.05$\\
      &&Ni&$14\pm7$&$7.0\pm2.2$\\
      &&Ni\footnotemark[$\S$]&$12\pm7$&$5.3\pm2.0$\\
      &$n_{\rm e}t$ (10$^{11}$~s~cm$^{-3}$)&&$3.5\pm0.4$&$20\pm3$\\
      &$EM$\footnotemark[$\dagger$] ($\times10^{12}$~cm$^{-5}$)&&$2.1\pm0.3$&$9.1\pm0.9$\\                    
      Gaussian (Al K$\alpha$) &Energy (keV)&&1.58$\pm$0.02&1.58 (fixed)\\
      &flux ($\times10^{-4}$~photons~s$^{-1}$~cm$^{-2}$)&&$2.0\pm0.7$&$3.3\pm2.3$\\      
      Gaussian (Fe K$\beta$)\footnotemark[$\|$] &Energy (keV)&&7.64 (fixed)&7.69 (fixed)\\
      &flux ($\times10^{-7}$~photons~s$^{-1}$~cm$^{-2}$)&&5.5 (fixed)&8.8 (fixed)\\      
                          \hline
$\chi ^2$/d.o.f.\footnotemark[$\S$] & & & $873/659=1.32$& $664/587=1.13$\\
\hline
\multicolumn{3}{@{}l@{}}{\hbox to 0pt{\parbox{180mm}{\footnotesize
\par\noindent
\footnotemark[$*$]Errors are at the 90\% confidence level.
\par\noindent
\footnotemark[$\dagger$]The emission measure in unit of $n_{\rm{e}}n_{\rm{H}}V/4\pi d^{2}$, where $n_{\rm{e}}$, $n_{\rm{H}}$, $V$ and $d$ are the electron density,\\
\ \  the hydrogen density, the emitting volume and the distance to the source, respectively.
\par\noindent
\footnotemark[$\ddagger$]The value is derived from the flux of Gaussian (Al K$\alpha$) (see section \ref{SNR_fit}).
\par\noindent
\footnotemark[$\S$]The value obtained from the revised three-component model.
\par\noindent
\footnotemark[$\|$]The Gaussian added in the case of the revised three-component model. Missing Fe K$\beta$ lines \\ 
\ \ except for H-like and He-like states for the NEI component are taken into account. 
}\hss}}
    \end{tabular}
     \end{center}
\end{table*}

\subsubsection*{G349.7$+$0.2}

As is shown in figure~\ref{fig:fit}, the spectrum of G349.7$+$0.2 is very similar to that of G350.1$-$0.3.
We, therefore, follow the same fitting process as is given above.
In all the fitting process of G349.7$+$0.2, we also include the CCO candidate, CXOU\,J171801.0$-$372617, as the power-law with $\Gamma =2.5$ and the unabsorbed 0.5--10.0~keV flux of $2.2\times10^{-13}$~erg~s$^{-1}$~cm$^{-2}$ \citep{Lazendic2005}.
We find  in figure~\ref{fig:fit}c that the NEI model fit is unacceptable with significant data residuals near at Fe K ($> 6.5$~keV), and the lower energy band ($< 3$~keV). 
Thus, we try the three-component model, which is the same as G350.1$-$0.3, and obtain an acceptable $\chi^2$/d.o.f. of $664/587=1.13$.
The best-fit model and its parameters are given in figure~\ref{tab:para} and table~\ref{tab:para}, respectively.
The abundance of Al and the revised abundance of Ni are estimated with the same method as in the case of G350.1$-$0.3. 
These are $0.6\pm0.4$~solar and $5.3\pm2.0$~solar for Al and Ni, respectively.

\section{Discussion}

The X-ray spectra of the two SNRs are well explained by the two plasma model; a high-temperature in NEI and a low-temperature in CIE.
For both the SNRs, we obtain the abundances of many heavy elements in the high-temperature NEI plasma.
The most important discovery is the detection of Ni with the extreme overabundance ($12\pm7$~solar for G350.1$-$0.3 and $5.3\pm2.0$~solar for G349.7$+$0.2).
We also find Al for the first time from G350.1$-$0.3 and G349.7$+$0.2, the second detection after G344.7$-$0.1 \citep{Yamaguchi2012}.

\subsection{Origin of the Plasmas}

\subsubsection{G350.1$-$0.3}\label{G350_origin}
\begin{figure}[tb]
\begin{center}
\FigureFile(80mm,70mm){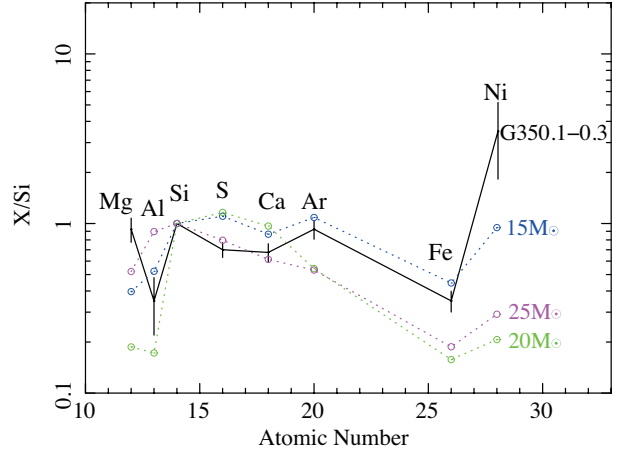}
\caption{Metal abundances in the ejecta of G350.1$-$0.3 relative to Si as a function of atomic number.
The dotted lines represent CC models with main sequence masses of 15~\MO, 20~\MO \ and 25~\MO\ \citep{Woosley1995}.}
\label{fig:pattern}
\end{center}
\end{figure}

Since the low-temperature component for G350.1$-$0.3 is in CIE with the solar abundances, it would be ISM heated by a blast wave.
The high-temperature NEI component for the SNR has high metal abundances of 1.4--12~solar. Therefore, it is likely the ejecta recently heated-up by a reverse shock.
In figure~\ref{fig:pattern}, we show metal abundances in the ejecta relative to Si for G350.1$-$0.3. 
The determination of Mg and Al abundances in the ejecta may be significantly affected by the high inferred Ni abundance, because Ni L-shell lines become important near Mg K$\alpha$ and Al K$\alpha$ line energies. Also the quality of atomic data for Ni lines in the NEI models is not good enough. 
Thus, the quoted errors on the Mg and Al abundances may be larger than the pure statistical error.
Taking into account of possible larger errors in Mg and Al than those given in figure~\ref{fig:pattern}, the abundance patterns roughly agree with those of the CC-SN model with a progenitor mass between 15--25~\MO \ \citep{Woosley1995}.

\subsubsection{G349.7$+$0.2}\label{G349_origin}
\begin{figure}[tb]
\begin{center}
\FigureFile(80mm,70mm){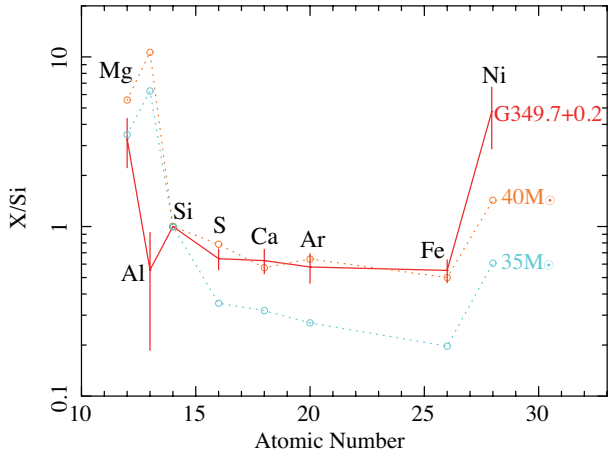}
\caption{Metal abundances in the ejecta of G349.7$+$0.2 relative to Si as a function of atomic number.
The dotted lines represent CC models with main sequence masses of 35~\MO \ and 40~\MO \ \citep{Woosley1995}.}
\label{fig:pattern2}
\end{center}
\end{figure}

Like G350.1$-$0.3, the low-temperature component for G349.7$+$0.2 is in CIE with the solar abundances.
Thus, this component is also likely ISM heated by a blast wave.
For the high-temperature component, Mg and Ni abundances are much higher than the solar values, suggesting an ejecta component. 
However, as we noted in \ref{G350_origin}, the Mg abundance would have a larger error.
Hence, compared to the case of G350.1$-$0.3, it would be less convincing that the high-temperature component of G349.7$+$0.2 is also an ejecta origin. 
Still, we show the abundance pattern of G349.7$+$0.2 in figure~\ref{fig:pattern2} comparing those of the CC-SN model with a progenitor mass between 35--40~\MO  \citep{Woosley1995}.
From figure~\ref{fig:pattern2}, we see the abundance pattern of the SNR roughly agrees with that of a progenitor mass of $\sim$35--40~\MO.
The metal abundances other than Mg and Ni are 1~solar or slightly smaller, which may conflict with the initial assumption of the ejecta origin.

The abundances are determined by fixing the abundances for lighter elements than Mg, namely He, C, N, O and Ne to be 1 solar since He--Ne do not appear as emission lines in the relevant energy band of $<1.2$~keV.
If the plasma is really due to the ejecta of 35--40~\MO\ star, the abundance of He--Ne should be far larger than 1 solar and the bremsstrahlung is largely dominated by the enhanced He--Ne.
We, hence, assume the abundances of these light elements following the results of \citet{Woosley1995}, and re-fit the spectra.
The resultant abundances of Mg--Ni become 4.5--5.5 times of the initial values, or larger than 1~solar, which supports the ejecta origin. 
The abundance ratios relative to Si are not changed from that of the original data given in figure~\ref{fig:pattern2}. 

\subsection{Ni Over-Abundance of G350.1$-$0.3 and G349.7$+$0.2}
The observed abundance of Ni are far higher than that of Fe for both the SNRs.
We note that the large abundances of Ni are not due to an estimation error of the NXB, which
exhibits a strong neutral Ni K$\alpha$ line at 7.47~keV. 
We estimate for the case of FI CCDs (XIS0 + XIS3), for simplicity. 
Neutral Ni K$\alpha$ line flux in the NXB is $3.1(\pm 0.1)\times10^{-6}$~photons~s$^{-1}$~cm$^{-2}$, 
which is almost comparable to the He-like Ni K$\alpha$ + the Fe K$\beta$ lines flux 
in the NXB-subtracted spectra of the SNRs($\sim3.5\times10^{-6}$~photons~s$^{-1}$~cm$^{-2}$).
Since the typical ambiguity of the NXB subtraction is at most 5\% \citep{Tawa2008}, 
a contamination of this line to the derived flux of the He-like Ni K$\alpha$ + Fe K$\beta$ lines would be less than a few \% . 
Furthermore, with the good energy resolution of Suzaku, we separately detect the He-like Ni K$\alpha$ + the Fe K$\beta$ lines 
at 7.7~keV from the neutral Ni K$\alpha$ line at 7.47~keV. 

\begin{figure}[tb]
\begin{center}
\FigureFile(85mm,85mm){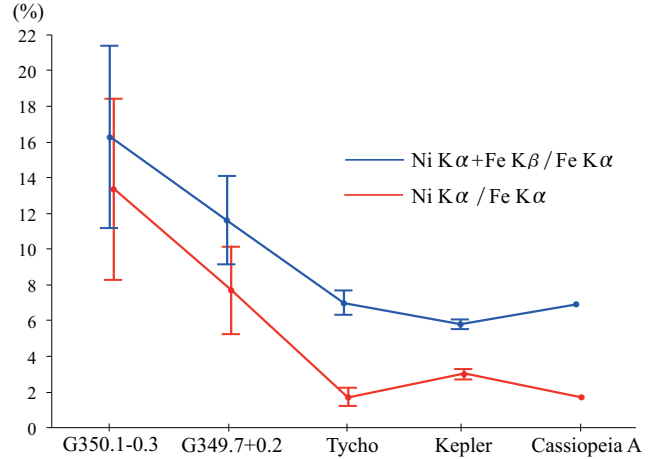}
\caption{Comparison with the flux ratio of K$\alpha$ lines between Fe and Ni. 
Blue points are those between Ni K$\alpha$ plus Fe K$\beta$ and Fe K$\alpha$ while red points are only for Ni K$\alpha$ and Fe K$\alpha$.
The errors are at the 1$\sigma$ level. 
Tycho and Kepler data are derived from \citet{Yamaguchi2014} and \citet{Park2013}, respectively.
The errors are not provided for Cassiopeia A \citep{Maeda2009}.}
\label{SNR_comparison}
\end{center}
\end{figure}

The high ratio of $Z_{\rm Ni}/Z_{\rm Fe}~\sim8$ is not found from any other SNRs.
In figure~\ref{SNR_comparison}, we compare simply the flux ratio of K$\alpha$ line of Fe and Ni for G350.1$-$0.3, G349.7$+$0.2, 
Tycho \citep{Yamaguchi2014}, Kepler \citep{Park2013}, and Cassiopeia A \citep{Maeda2009}.
For G350.1$-$0.3, G349.7$+$0.2, and Cassiopeia A, only the sum of Ni K$\alpha$ and Fe K$\beta$ are available.
We therefore estimated the Fe K$\beta$ flux by referring to \citet{Yamaguchi2014}, and obtain the Ni K$\alpha$ flux separately.
Since the atomic numbers of Fe and Ni are nearly the same, the flux ratio of K$\alpha$ line of Ni and Fe is approximately equal, or slightly smaller (due to a smaller ionization/excitation cross section of Ni than those of Fe) than the abundance ratio.  
In the solar abundance, the abundance ratio Ni/Fe is $\sim$ 4\%. 
The flux ratios of K$\alpha$ line of Ni and Fe for Tycho, Kepler and Cassiopeia A are slightly smaller than $\sim$ 4\%, but those of G350.1$-$0.3 and G349.7$+$0.2 are larger than $\sim$ 4\%, indicating that the abundance ratio ($Z_{\rm Ni}/Z_{\rm Fe}$) is larger than 1 (in solar unit) in these SNRs.
Thus Ni-overabundance for G350.1$-$0.3 and G349.7$+$0.2 can be suggested even before the spectral fitting. 

The high ratio $Z_{\rm Ni}/Z_{\rm Fe}~\sim8$  is not predicted by the theoretical model by \citet{Woosley1995}. 
The observed high ratio can be explained if a significant fraction of Ni was ejected from the core region possibly due to an asymmetric explosion.
In fact, \citet{Maeda2007} reported that a large amount of Ni is ejected from the core of SN 2006aj as a result of an asymmetric explosion.
G350.1$-$0.3 and G349.7$+$0.2 has the morphology away from symmetry. 
Previous researches claimed that the surrounding molecular gas caused the peculiar morphologies.
Instead, we propose that asymmetric explosions made such morphologies.

\subsection{$N_{\rm H}$, Distance and Ejecta Mass}

The absorption ($N_{\rm H}$) of compact sources in the Galactic inner plane would be affected by the dust scattering effect. 
The observed radius of the dust scattering halo (which includes 90\% of the total flux) is $\sim \timeform{40"}$ \citep{Xiang2007} for 4U\,1624$-$49, an X-ray binary located at or behind the Galactic ridge with large absorption of $N_{\rm H} \sim 8\times10^{22}$~cm$^{-2}$ \citep{Smale2001}. 
Although the source sizes of G350.1$-$0.3 and G349.7$+$0.2 are larger than this radius, we still examine the dust scattering effect. 
We made the spectra of the SNRs from areas larger by \timeform{180"} in radius than those in figure~\ref{img} (solid lines) (e.g. for G349+0.2, the radius of the larger area is \timeform{320"}, while the original source area is \timeform{140"} radius.).
The best-fit $N_{\rm H}$ of the spectra from the larger areas are $3.2(\pm 0.1)$ and $6.3(\pm 0.2)\times 10^{22}$~cm$^{-2}$ for G350.1$-$0.3 and G349.7$+$0.2, respectively, which are consistent with those given in table~\ref{tab:para}.
Therefore, the dust scattering effect is not significant in the $N_{\rm H}$ estimation for these SNRs.

For the distance estimation, we assume that the interstellar gas density is proportional to the stellar density of the Galactic disk given by \citet{Kent1991}.
The ratio of the X-ray absorption column density $N_{\rm H}$ between the SNRs and nearby GRXE are 1.1 and 1.6 for G350.1$-$0.3 and G349.7$+$0.2, respectively (see tables \ref{tab:bg} and \ref{tab:para}).  
Integrating the gas density along the line of sight, we search for the distance, where the integrated gas density becomes to $N_{\rm H}$ (at 8.5 kpc) $\times$ $N_{\rm H}$ ratio (1.1 for G350.1$-$0.3 and 1.6 for G349.7$+$0.2).
Here we assume $N\rm_H$ of the GRXE is that of the midpoint of the Galactic ridge along the line of sight (8.5~kpc). Then the distances are estimated to be  $8.9\pm0.3$~kpc and $11.9\pm0.4$~kpc for G350.1$-$0.3 and G349.7$+$0.2, respectively.

Since the stellar density model \citep{Kent1991} does not include local enhancement of interstellar media (e.g. the 3~kpc arms; \cite{Dame2008}), we make the IR extinction curves by \citet{Chen2013}\footnote{The on-line data of \citet{Chen2013} are limited in the distance of below 10~kpc and in the longitude below \timeform{10D} from the Galactic center. Dr. Chen kindly provided us with the data near at $l$=\timeform{350D} up to distance of $\sim14$~kpc.}, and re-estimate the distance with the same method as described above. 
Then the re-estimated distance of G350.1$-$0.3 is $9.4\pm0.4$~kpc, consistent with that taken from the stellar density model ($8.9\pm0.3$~kpc).
No IR extinction curve is available at the position of G349.7$+$0.2. 
We therefore use one of the nearby data at $l\sim\timeform{350D}$, and obtain a distance of $12.7\pm0.6$~kpc, which is also consistent with that derived from the stellar density model ($11.9\pm0.4$~kpc).
However, the near-by data show significant spatial variations, and are different from \citet{Marshall2006}. 
We estimate the distance variation using these data, and found the variations to be $\sim$2--3~kpc. 
Thus we regard the systematic distance error for G349.7$+$0.2 using the current IR extinction data is $\sim$2--3~kpc.

The variation of the $N_{\rm H}$ obtained by the X-ray observations in the $\timeform{2D}< |l| < \timeform{10D}$, 
$ \timeform{-0.5D} < b < \timeform{0.5D}$ region is less than 30\% (90\% error) (H. Uchiyama, private communication).
This would be another source of the distance uncertainty.
Taking into account of all these possible systematic errors, we adopt the distances of G350.1$-$0.3 and G349.7$+$0.2 to be 9$\pm$3 and 12$\pm$5 kpc, respectively.
The distance of G350.1$-$0.3 is consistent with, while that G349.7$+$0.2 is smaller than those of the previous reports \citep{Gaensler2008,Caswell1975,Frail1996}.

We will estimate the ejecta masses for both the SNRs as below.
As we mentioned in \ref{G349_origin}, we should deal with lighter elements which do not appear in the relevant energy band of $>$ 1.2~keV to estimate the physical condition of the plasma such as the emission measure ($EM$).
Since the ejecta abundances are similar to those of 15--25~\MO\ and 35--40~\MO\ progenitor stars for G350.1$-$0.3 and G349.7$+$0.2, respectively, we assume that the abundances of elements lighter than Mg in the NEI component (ejecta) are those of the CC-SN model of a 20~\MO\ and 40~\MO\ progenitor \citep{Woosley1995} for G350.1$-$0.3 and G349.7$+$0.2, respectively and re-fit the spectra.
As a result, the emission measures become $\sim$1/4 and $\sim$1/5, 5.4($\pm$0.5)$\times$10$^{11}$~cm$^{-5}$ and 1.7$(\pm$0.2)$\times$10$^{12}$~cm$^{-5}$ for G350.1$-$0.3 and G349.7$+$0.2, respectively.
We then take the ratio between the electron and atomic hydrogen densities to be $n_{\rm e}/n_{\rm H} =$1.6 and 1.7 and the number ratio of all the nucleons to hydrogen to be 2.1 and 2.5 in the 20~\MO\ and 40~\MO\ progenitor, respectively.
Assuming an oblate spheroid with major and minor radii of $\timeform{2.3'}$ and $\timeform{1.5'}$ for G350.1$-$0.3 \citep{Lovchinsky2011} and a sphere with a radius of $\timeform{1.2'}$ for G349.7$+$0.2 \citep{Lazendic2005}, the ejecta masses are estimated to be $\sim$13~$f^{1/2}~d^{5/2}_9$~\MO\ for G350.1$-$0.3 and $\sim$24~$f^{1/2}~d^{5/2}_{12}$~\MO\ for G349.7$+$0.2, where $f$ is a filling factor.
$d_9$ and $d_{12}$ are the distance parameters in unit of 9~kpc and 12~kpc, respectively.
These are roughly consistent with those estimated by the abundance patterns of 15--25~\MO\ and 35--40~\MO, respectively.

\section*{ADDED IN PROOF:}
After the submission of this draft, \citet{Tian2014} reported a revised kinematic distance of G349.7$+$0.2 to be about 11.5 kpc, which agrees well with our result.

\bigskip
The authors are grateful to Drs.~Hiroya Yamaguchi, Hideki Uchiyama, Keiichi Maeda and Nozomu Tominaga for valuable information and comments, and to Dr. Bingqiu Chen for the NIR extinction data used in distance estimation. 
The authors also thank all the Suzaku team members for their developing of the hardware and software, spacecraft operations, and instrument calibrations. 
S.N. and H.U. are supported by Japan Society for the Promotion of Science (JSPS) Research Fellowship for Young Scientists. 
This work is supported by JSPS Scientific Research grant numbers 24740123 (M.N.), 20340043, 23340047 and 25109004 (T.G.T.), 20600406 (T.T.), 23000004 and 24540229 (K.K.).

\end{document}